\documentclass[10pt, format=manuscript,nonacm]{acmart}

\AtBeginDocument{%
  \providecommand\BibTeX{{%
    \normalfont B\kern-0.5em{\scshape i\kern-0.25em b}\kern-0.8em\TeX}}}

\usepackage{array}
\usepackage{booktabs}
\usepackage{hyperref}
\usepackage{graphicx}
\usepackage{threeparttable}
\usepackage{setspace}
\usepackage{amsmath}

\usepackage{amssymb}
\usepackage[marginal]{footmisc}
\makeatletter
\def\UrlAlphabet{%
  \do\a\do\b\do\c\do\d\do\e\do\f\do\g\do\h\do\i\do\j%
  \do\k\do\l\do\m\do\n\do\o\do\p\do\q\do\r\do\s\do\t%
  \do\u\do\v\do\w\do\x\do\y\do\z\do\A\do\B\do\C\do\D%
  \do\E\do\F\do\G\do\H\do\I\do\J\do\K\do\L\do\M\do\N%
  \do\O\do\P\do\Q\do\R\do\S\do\T\do\U\do\V\do\W\do\X%
  \do\Y\do\Z}
\def\UrlDigits{\do\1\do\2\do\3\do\4\do\5\do\6\do\7\do\8\do\9\do\0}
\g@addto@macro{\UrlBreaks}{\UrlOrds}
\g@addto@macro{\UrlBreaks}{\UrlAlphabet}
\g@addto@macro{\UrlBreaks}{\UrlDigits}

\begin{document}

\title{How to optimize an academic team when the outlier member is leaving?}

\author{Shuo Yu}
\affiliation{%
\department{School of Software}
\institution{Dalian University of Technology}
\city{Dalian}
\postcode{116620}
\country{China}}

\author{Jiaying Liu}
\affiliation{%
\department{School of Software}
\institution{Dalian University of Technology}
\city{Dalian}
\postcode{116620}
\country{China}}

\author{Feng Xia}
\affiliation{%
\department{School of Engineering, IT and Physical Sciences}
\institution{Federation University Australia}
\city{Ballarat}
\postcode{VIC 3353}
\country{Australia}}

\author{Haoran Wei}
\affiliation{%
\department{School of Software}
\institution{Dalian University of Technology}
\city{Dalian}
\postcode{116620}
\country{China}}

\author{Hanghang Tong}
\affiliation{%
\institution{University of Illinois at Urbana-Champaign}
}


\begin{abstract}
 An academic team is a highly-cohesive collaboration group of scholars, which has been recognized as an effective way to improve scientific output in terms of both quality and quantity. However, the high staff turnover brings about a series of problems that may have negative influence on team performance. To address this challenge, we first detect the tendency of the member who may potentially leave. Here the outlierness is defined with respect to familiarity, which is quantified by using collaboration intensity. It is assumed that if a team member has a higher familiarity with scholars outside the team, then this member might probably leave the team. To minimize the influence caused by the leaving of such an outlier member, we propose an optimization solution to find a proper candidate who can replace the outlier member. Based on random walk with graph kernel, our solution involves familiarity matching, skill matching, as well as structure matching. The proposed approach proves to be effective and outperforms existing methods when applied to computer science academic teams.
\end{abstract}




\maketitle

Academic teams are regarded as one of the most fundamental research patterns, which have proved to be effective in improving scientific output as well as research quality. There are many studies that focused on how to optimize academic teams in various scenarios such as team member replacement, team expansion, as well as team shrinkage. A branch of science entitled Science of Scientific Team Science has emerged with the goal to make the most of academic teams. Under many circumstances, policy makers, project leaders, as well as administrators need to make certain personnel adjustments based on, e.g., members’ contributions in teamwork. Detecting an outlier member and optimizing the corresponding academic team have become critical problems in many areas including management science, organization science, social science, etc.

It has been proved that members with stable collaboration relationships can improve team performance~\cite{lee2015effects}. However, academic teams often suffer from a high turnover due to many reasons.

Stable collaboration pattern helps with higher quality output in academic teams. However, academic teams are always facing with the problem of departure of team members. In this work

A better way for this problem is to prepare for choosing an alternative for this outlier member. To better identify the outlier member, network science technologies can be employed. Network science has significantly changed academic team research, and has also proved to be an effective tool to detect outliers~\cite{tiropanis2015network,wachowicz2016finding}. Meanwhile, networked data might be more useful than other data forms in the non-IID (independently identically distribution) context. Various scenarios such as team member replacement, team expansion, as well as team shrinkage have been studied by using network-based methods. In many cases, an academic team can be converted to a subgraph which is embedded in a large-scale academic social network. The underlying mechanism of team optimization with networks has been explored in the literature~\cite{li2015replacing}. Some relevant algorithms assemble optimal academic teams by employing random walk graph kernel, encoding academic skills to minimize communication costs within the team~\cite{li2015replacing}.

Most studies in this direction deal with team formation or specific team enhancement~\cite{butchibabu2016implicit,kim2017makes}. They often consider the matching of members' skills and the completion of tasks to find the optimal solution with the lowest communication costs~\cite{kargar2012efficient,ashenagar2015team,chen2017analysis}. However, academic collaboration through teamwork is a typical social behavior~\cite{fehr2018normative}. That is, the closeness among members is critical in detecting an outlier and is therefore also important in enhancing teamwork~\cite{hinds2000choosing}. Here we implement a familiarity model that seeks to quantify the collaboration closeness among team members. Outlier members can be easily detected through such a proper implementation.

\begin{figure}
  \centerline{\includegraphics[width=18.5pc]{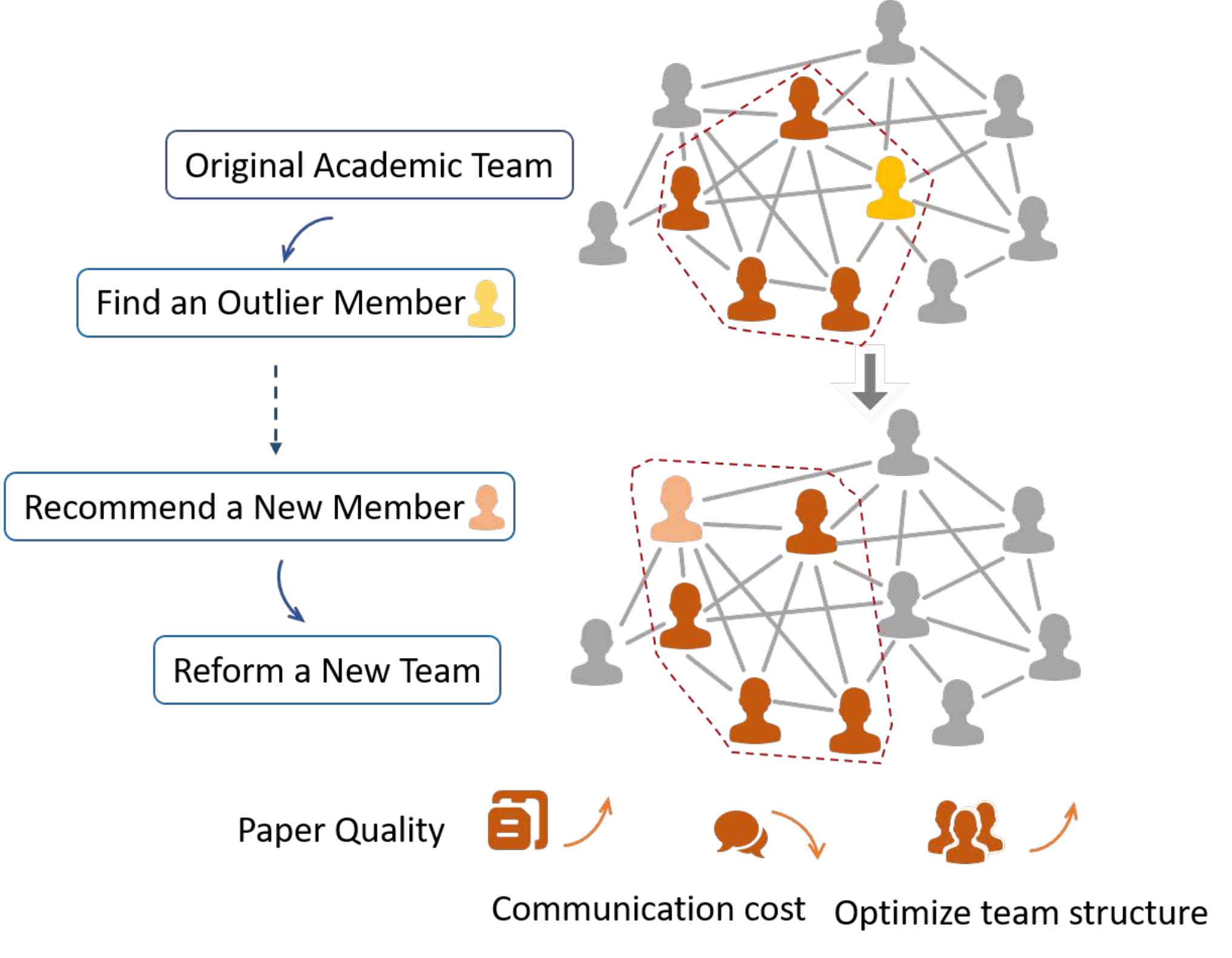}}
  \caption{The outlier member replacement problem.}
  \label{fig:problem}
\end{figure}

Network sub-structures such as components, graphlets, and motifs can be used to effectively describe specific real-world patterns~\cite{benson2016higher,lin2016network}. Among them, network motifs have proved to be the most proper sub-structure in describing meso-level human behavior~\cite{milo2002network}. It has also been verified that members within a same academic team sub-unit possess a relatively higher collaboration intensity than those outside the unit~\cite{yu2017team}. Such sub-structures can be explored to capture the multi-variate relations among members, and can thus also be used to evaluate the degree of familiarity. In this paper, the principle of network motif is used to evaluate the familiarity among team members. According to our statistics, members possessing a lower degree of familiarity generally possess a higher probability of leaving, and are defined as outlier members. The problem is explicitly defined in Figure~\ref{fig:problem}. By considering both pairwise and higher-order familiarity, we can effectively identify outlier members within a given academic team. We also propose an enhanced team member replacement algorithm called OMR (Outlier Member Replacement). In OMR, we take familiarity matching, skill matching, and structure matching into consideration. Our proposed algorithm outperforms in optimizing academic teams. In comparison to the baseline methods, experimental results show that OMR performs better than TFP in both the quality of team output and the communication cost within the team. Moreover, our proposed OMR can be applied to other similar scenarios such as teams in business and industry,  with a proper adaptability definition of familiarity based on former cooperation relationships.

\begin{table*}[ht!]
  \centering
  \caption{Notations and Symbols}
  \begin{tabular}{p{3cm}<{\centering}p{7cm}<{\centering}}
    \hline
    Notation & Description \\
    \hline
    $||F||_1$ & Pairwise familiarity \\
    $||F||_n$ & Higher-order familiarity \\
    $PairwiseCol_{ij}$ & Pairwise collaboration relationship between $i$ and $j$ \\
    $MultiCol_{ij}$ & Higher-order collaboration relationship among $i$ and $j$ \\
    $T, T_p, T'$ & Original team, team after member $p$ left, team with alternate member \\
    $Out_i$ & Outlier degree between $i$ and $i$'s team \\
    $p$ & The member who is about to leave the team \\
    $cand$ & Candidate member \\
    $I$ & Identity matrix \\
    $\mu$ & Decay parameter \\
    $F_{T_p}^p$ & Familiarity matrix of member $p$ and team $T_p$ \\
    $S_T^j$ & The skill matrix of $T$ \\
    $G,V,C,S$ & Collaboration network, vertices set, edge set, skill set\\
    \hline
  \end{tabular}
  \label{tab:n}
\end{table*}
\section{FAMILIARITY AND OUTLIER MEMBER DETECTION}

Two forms of familiarity are defined based on their network sub-structures. To better explain the method we proposed, we summarize the notations in Table~\ref{tab:n}. Specifically, we define pairwise familiarity from bipartisan collaboration relationships and higher-order familiarity from multi-variate collaboration relationships.

Pairwise familiarity is defined in Eq.~\ref{eq:pf}.

\begin{equation}
  ||F||_1(i, T)=\sum_{j\in T}PairwiseCol_{ij}.
\label{eq:pf}
\end{equation}
Wherein, $PairwiseCol_{ij}$ reflects the collaboration relationship between member $i\in T$ and member $j$. If $i$ and $j$ have collaborated before, then $PairwiseCol_{ij}=1$. Otherwise, $PairwiseCol_{ij}=0$.

Higher-order familiarity is defined in Eq.~\ref{eq:hf}.

\begin{equation}
||F||_{n}(i,T)=\sum_{j\in T}MultiCol_{ij}.
\label{eq:hf}
\end{equation}
Wherein, $MultiCol_{ij}$ reflects whether member $i \in T$ and member $j$ are involved in a certain multivariate relationship~\cite{xu2020multivariate}. If both $i$ and $j$ are involved in a specific multivariate relationship, then $MultiCol_{ij}=1$. Otherwise, $MultiCol_{ij}=0$.

Pairwise collaboration relationships are generally modeled using edges in academic networks, while multivariate collaboration relationships are generally modeled using network motifs.

The extent of a member being an outlier, is defined as the ratio of one’s familiarity within a team, as it compares to all of his/her collaboration relationships. In other words, if a member bares a relatively lower familiarity for a given period, then he/she is regarded as an outlier who is probably going to leave the team.

\begin{equation}
Out_{i}=\frac{||F||_{n}(i,T)}{MultiCol_{i}}.
\end{equation}

\section{TEAM OPTIMIZATION}

Teams can be optimized in various ways, including member replacement, team refinement, team expansion, team shrinkage, among others. We discuss a specific situation where the objective is to optimize the team when the outlier member is leaving. In order to improve the efficacy of familiarity evaluation, team member replacement is utilized to optimize a specific academic team. The general purpose of team member replacement is to find a similar member $m_{rep}$ to replace the outlier $m_{out}$. In an attempt to maintain or even improve team performance, the replacement $m_{rep}$ is expected to possess a similar skill-set with $m_{out}$. It is also significant to ensure that the new member has a functional collaboration basis with the original members. This paper defines familiarity by evaluating the basis for collaboration. In other words, the new member is expected to maintain a similar collaboration structure as the outlier member. Therefore, the following three matching targets exist.

In general, skill matching is the notion that the new member is expected to have a similar skill set as the outlier member. This is particularly important for teams in which that specific skill is integral and vital. Structure matching means that the new member is expected to maintain the network structure of the team, so that it does not collapse in the absence of outlier. Familiarity matching is based on the idea that the new member is expected to have a high degree of familiarity with the rest of the team members so as to maintain a similar extent of communication costs.

To satisfy the above mentioned demands, the similarity between two teams must be evaluated. Herein, a graph kernel approach is utilized. Most of the proposed graph kernel methods are based on the similarity between two input graphs~\cite{li2015replacing}. Consequently, this paper utilizes a graph kernel to capture the extent of skill matching, structure matching, and familiarity matching. However, it is important to note that graph kernels are not merely a combination of several features. The graph kernel proposed in this work is defined below.

For a given team $T_p$ and complete scientific collaboration network $G(V,C,S)$, we recommend $k$ candidates from $G$ based on structure matching, skills matching, and high familiarity of candidates. Structure matching means that the new candidate should possess a similar topological graph structure to the member who is departing. Skill matching means that the new candidate should possess similar skills to that of the departing member. As for familiarity, it is expected that the new candidate should be more familiar with the remaining team members. We employ a graph kernel to define the OMR score as follows to evaluate the effectiveness of team optimization.

\begin{equation}
\label{eq:omr}
\begin{aligned}
& OMR(G,T,p,cand)=\mathop{\arg\max}_{i\in cand}x_{1}^{\intercal} \\
&\left(I-\mu G_{T_{p}}\right)^{-1} \sum_{j=0}^{n_{S}}\left(F_{T_{p}}^{p} S_{T}^{j} \otimes F_{T_{p}}^{i} S_{T^{\prime}}^{j}\right) x_{2}
\end{aligned}
\end{equation}

Wherein, $x_1$ and $x_2$ represent start vector and end vector, respectively. $I$ is the identity matrix and $G_{T_P}$ is Kronecker product of original team and optimized team, i.e., $G_{T_P}=G_T\otimes G_{T'}$. $\mu$ refers to the decay parameter. $F_{{T}_p}^{p}, F_{{T}_p}^{i}$ are familiarity matrices of original team and the optimized team with candidate member in it, respectively.

\begin{figure*}
  \centerline{\includegraphics[width=\linewidth]{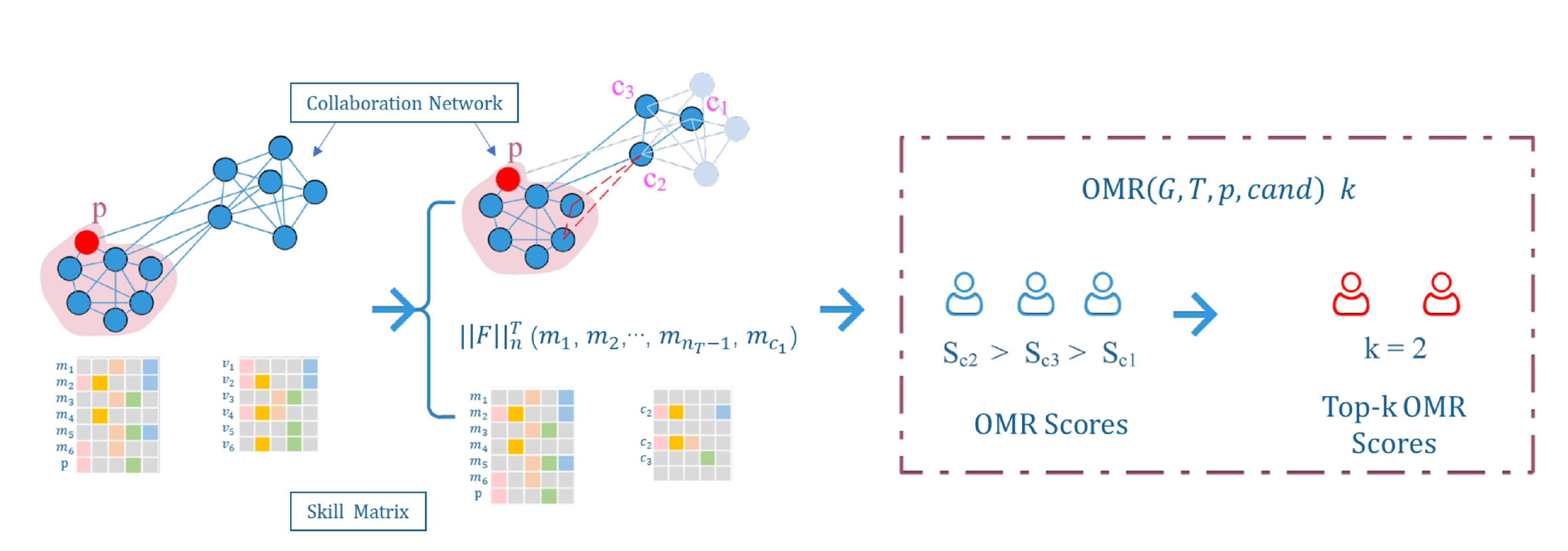}}
  \caption{The overall framework of OMR.}
  \label{fig2}
\end{figure*}

The process of outlier member replacement is perhaps one of the most tenable solutions for team optimization when an outlier member has left. A list of remaining candidates and their direct relationships can be constructed, by considering the collaboration relationships of all members within the network. By calculating the degree of familiarity amongst all candidates, such familiarity score is used as the diagonal values of an input matrix. General scores for all possible teams for each candidate are obtained using Eq.~\ref{eq:omr}. Higher scores of familiarity indicate greater degrees of appreciation for said candidate. The overall process of OMR is shown in Figure~\ref{fig2}.

The computational complexity of OMR is then analyzed. Details are listed in Table~\ref{tab:cc}. To be specified, we only list pairwise familiarity calculation here because the computational complexity of higher-order familiarity calculation depends on the order of employed motif. In Table~\ref{tab:cc}, $c$ refers to the number of candidates, $t$ is the number of team members, and $r$ is the total number of skills.
\begin{table}
\centering
\caption{Computational complexity of OMR}
\begin{tabular}{cc}
  \hline
  Step & Computational Complexity \\
  \hline
  Pairwise familiarity calculation & $O(t^2)$ \\
  Outlier member detection & $O(t)$ \\
  Edge weights calculation & $O(|C|)$ \\
  Skill similarity calculation & $O(r\times t)$ \\
  OMR score calculation & $O(ct^2r^2)$ \\
  \hline
\end{tabular}
\label{tab:cc}
\end{table}

\section{TEAM OPTIMIZATION IN REAL-WORLD NETWORKS}

We apply our method to CiteSeerX dataset~\footnote{\url{http://citeseer.ist.psu.edu/}} and Microsoft Academic Graph (MAG)~\footnote{\url{https://www.microsoft.com/en-us/research/project/microsoft-academic-graph/}}. CiteSeerX contains abundant computer and information science literatures, metadata, algorithms, as well as academic services. MAG is a heterogeneous graph containing scientific records and relative information, which covers multiple disciplines.
We respectively extract data in computer science discipline. We employ 15,681 scholars and 42,999 collaboration relations are selected in CiteSeerX dataset. As for MAG dataset, 252,439 scholars and 436,905 collaboration relationships are used. Details of two datasets are shown in Table~\ref{tab:d}.

Both of two data sets are divided into two sets, i.e., standard set and testing set. We artificially extract the outlier members from who leave their teams during the year 2013 to 2015. This is regarded as standard set. And then we use data from 2005 to 2012 (i.e., testing set) to recognize outlier members and verify our identification results in the standard set. In CiteSeerX dataset, the whole running time of OMR\_H and OMR\_P are 7.036 and 184.116 seconds, respectively. Since MAG dataset is in larger scale than CiteSeerX, OMR\_H and OMR\_P consume more time, i.e., 572.704 and 1339.808 seconds, respectively.

\begin{table}
\centering
\caption{Detail information of CiteSeerX and MAG datasets}
\begin{tabular}{ccc}
  \hline
  Properties & CiteSeerX & MAG \\
  \hline
  Time period & 2005-2015 & 2005-2015\\
  Number of nodes & 15,681 & 252,439 \\
  Number of edges & 252,439 & 436,905 \\
  Number of teams & 4,554 &  111,214 \\
  Number of motifs & 106,092 & 648,055\\
  \hline
\end{tabular}
\label{tab:d}
\end{table}

Using members and outliers together as filter condition, we firstly identify 791 outlier members from 939 academic co-author teams in CiteSeerX dataset and 3,228 outlier members from 5,413 teams in MAG dataset. All of the outliers have been verified to leave the team relatively soon (generally within 2 years). Then we use the above two mentioned

After detecting outliers, we tabulate the statistics of members who leave their original teams. OMR involves familiarity, skill, and structural features to recommend proper members based on their absences in academic teams. We also employ several baseline methods, i.e., TFP~\cite{yin2018social}. TFP considers both skill requirements and social connections in recommending candidates. The main idea of TFP is to find a candidate whose structural similarity and social connections are similar to the outlier member. It is proved to be an integrated approach to combine skill feature with structural feature.

Seven distinct approaches are implemented to optimize academic teams. We employ both pairwise familiarity and higher-order familiarity in OMR. More specifically, OMR\_H and OMR\_P refer to the proposed methods that employ higher-order and pairwise familiarity, respectively. TFP only consider skill requirements and structural features. The rest of comparison methods such as Kernel, High-order, Pairwise, and Skill, consider single character matching.

All of the above methods have been implemented and compared with average accuracy and sum distance. Average accuracy is calculated by the formula shown in the following equation. Wherein, $Q$ is the recommendation list and $Q_{real}$ is the set containing those who actually join the team. Accuracy ranges from 0 to 1.

\begin{equation}
Accuracy=\frac{|Q\cap Q_{real}|}{Q_{real}}
\end{equation}

We also employ average shortest path (shown in Eq.~\ref{eq:path}) and sum distance (shown in Eq.~\ref{eq:sum}) to evaluate the communication costs. A team with lower average shortest path refers that members within the team have lower communication costs. Meanwhile, sum distance is also employed to evaluate the distance between two members with different skills. A team with lower sum distance represents that members with different skills having lower communication costs.

\begin{equation}
AvgPath_T=\frac{2\sum_{i,j\in T}\min{PathLength_{ij}}}{n_T(n_T-1)}
\label{eq:path}
\end{equation}
Wherein, $\min PathLength_{ij}$ refers to the shortest path length between $i$ and $j$. $n_T$ is the number of members in team $T$.

\begin{equation}
Sumdistance_T=\sum_{i=1}^{n_s}\sum_{j=i+1}^{n_s}distance(v(s_i),v(s_j))
\label{eq:sum}
\end{equation}
Wherein, $v(s_i)$ refers to the scholar with skill $s_i$ and $n_s$ refers to the number of skills.

\begin{figure}
  \centerline{\includegraphics[width=18.5pc]{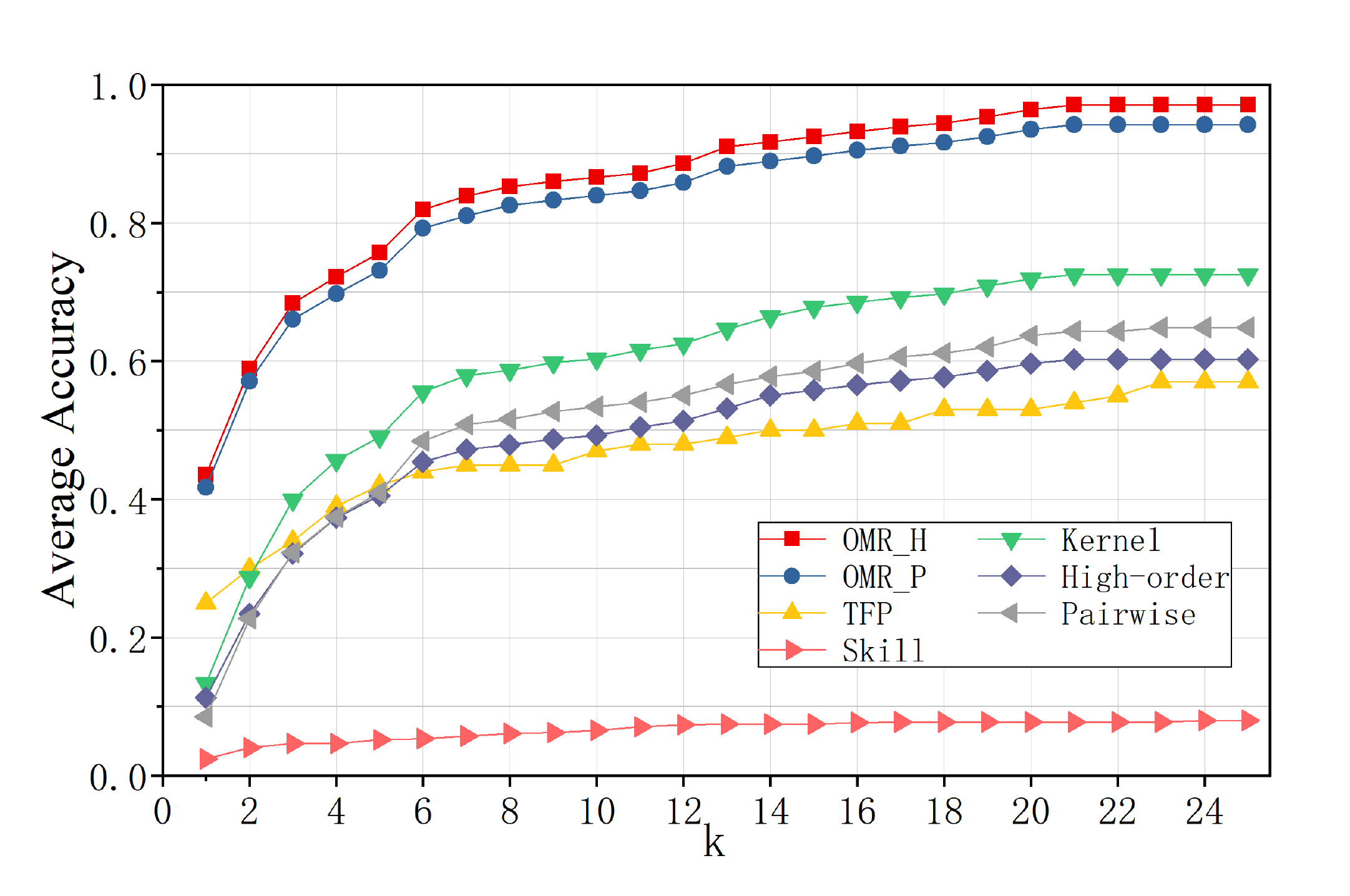}}
  \caption{Average accuracy of team optimization algorithms in CiteSeerX dataset.}
  \label{acite}
\end{figure}

\begin{figure}
  \centerline{\includegraphics[width=18.5pc]{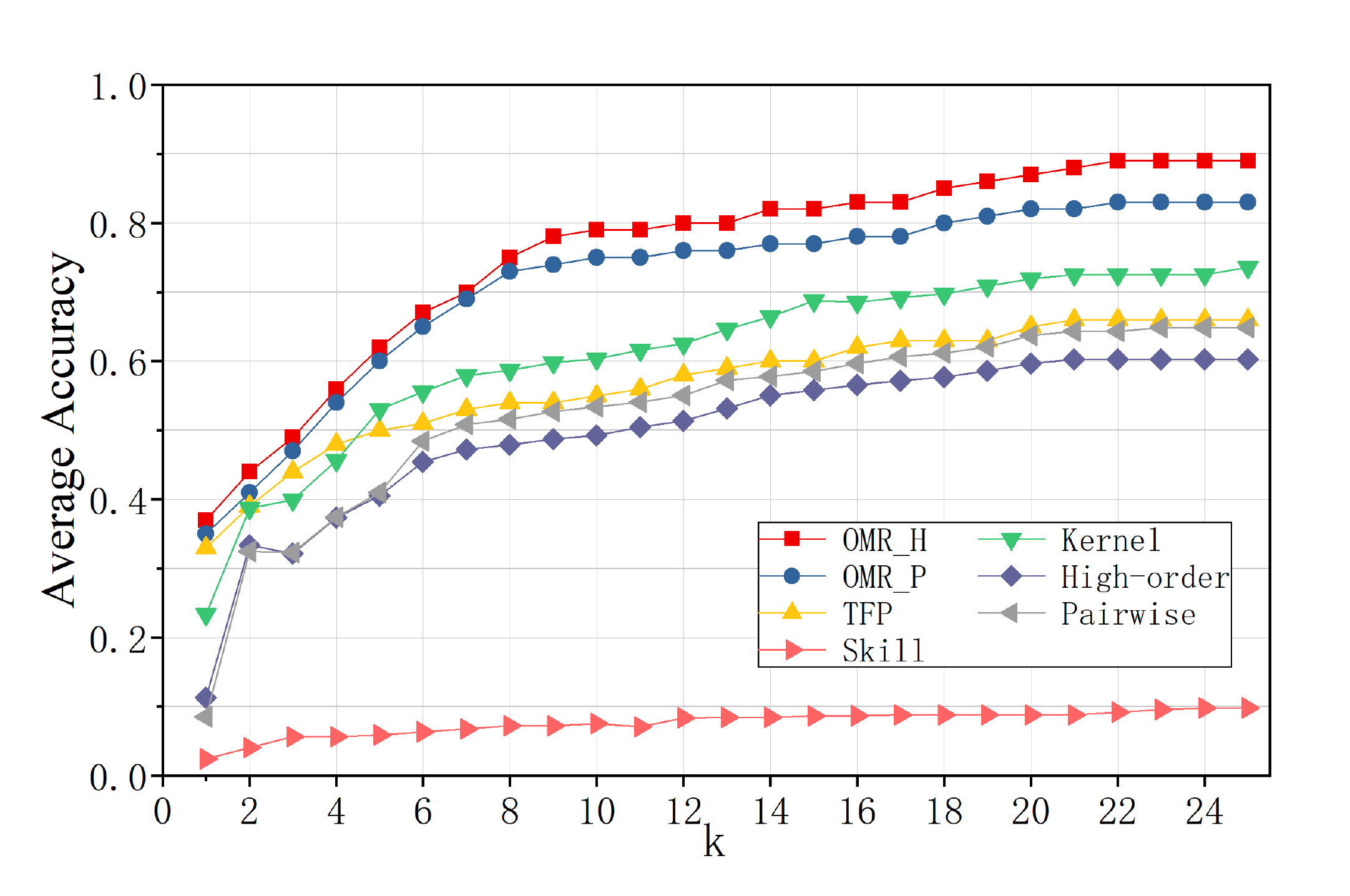}}
  \caption{Average accuracy of team optimization algorithms in MAG dataset.}
  \label{amag}
\end{figure}

Experimental results (shown in Figure~\ref{acite} and Figure~\ref{amag}) indicate that OMR outperforms others in terms of average accuracy. A higher-order of familiarity will lead to a higher accuracy. OMR with both higher-order and pairwise familiarity achieves significantly greater average accuracy than other methods, thereby yielding a better performance. Familiarity occupies an important position in optimizing academic teams. In contrast, the methods that employ matching of a single feature, perform worse, especially when such a feature is skills. This is perhaps because teamwork requires a highly-cohesive collaboration. Hence the consideration of familiarity is quite significant.

Moreover, a rigid approach that only considers skills, performs the worst. This does not indicate that skill is not important to team optimization. On the contrary, the precise skill set of a member plays a critical role in teamwork and is often regarded as the basis for team member replacement during optimization~\cite{kong2019skill}. If there is a skill mismatch, the replacement is essentially meaningless. Once the relevant skill set is matched, improved structural features and higher familiarity will ultimately lead to a more effective team.

The following considerations must be made when improving academic teams, including both internal consumption and external output. The external output of an academic team is often evaluated using technical or scientific metrics. When evaluating academic outputs such as papers, patents, and proposals, there exist some universal metrics such as citation counts and publication number. While internal consumption is usually quantified using communication costs. Communication costs are widely applied in various team-based problems~\cite{kargar2012efficient}. There also exist several ways to quantify communication costs. Among them, average shortest path length and sum distance are two of the commonly used metrics.

We specifically focus on the growth of paper citation counts. For those teams that replace their outlier members with higher-order familiarity, over 74.5\% of the teams in CiteSeerX, achieve higher citation counts. And that ratio in MAG is 80.7\%. The proportion is 71.6\% for CiteSeerX and 75.7\% for MAG if we employ pairwise familiarity. However, TFP only achieve 62.3\%, 60.3\%, 64.2\%, respectively. This indicates that our proposed method OMR leads to improved team performance.

The average shortest path length and sum distance are considered for internal consumption. According to our statistics, there are only 5\% of teams having 9 or more members in CiterSeerX and 6\% in MAG, which occupies our work focuses on teams of size 3 to 9 members. This range should be adjusted accordingly for other disciplines. Existing research specifically indicates how to select an appropriate team size, but that is outside the scope of this paper. In general, OMR achieves a lower average shortest path, when compared to the baseline methods.

OMR\_H achieves the lowest average shortest path, while OMR\_P has a similar performance to OMR\_H yet slightly worse than OMR\_H. In CiteSeerX dataset, OMR\_H reduces the average path by 7\%, 33\%, 32\%, 35\%, 10\%, 11\%, 14\% for teams of size 3 to 9, respectively. The baseline method, i.e., TFP aptly increases the average path length when team size is 3 to 9. As for MAG dataset, OMR\_H reduces the average path by 8\%, 12\%, 17\%, 24\%, 21\%, 8\%, 12\%, respectively. TFP only decreases the average path length when team size is 3, 4, and 7.

\begin{figure}
  \centerline{\includegraphics[width=18.5pc]{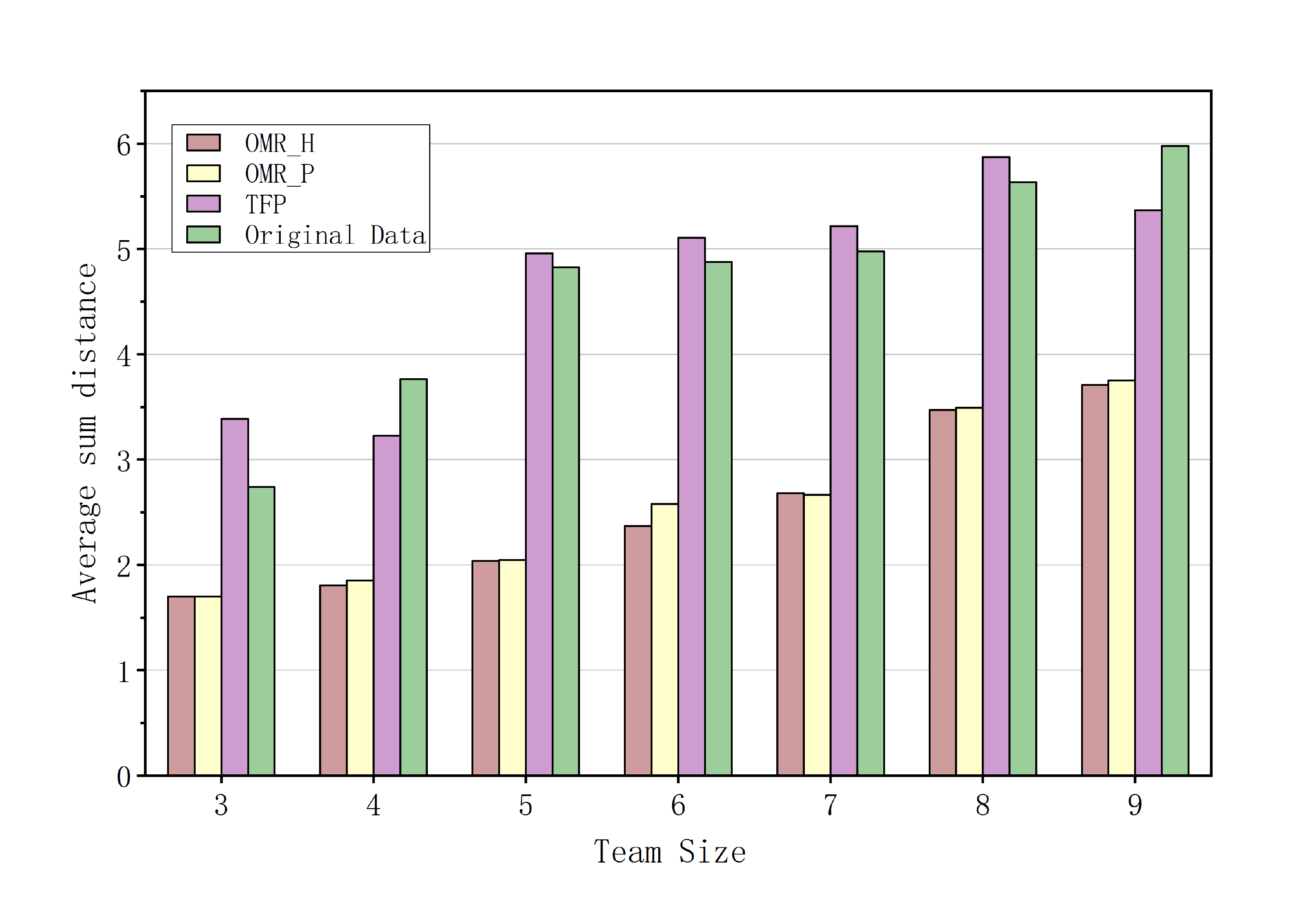}}
  \caption{Comparison results of sum distance in CiteSeerX dataset.}
  \label{fig:sum}
\end{figure}

\begin{figure}
  \centerline{\includegraphics[width=18.5pc]{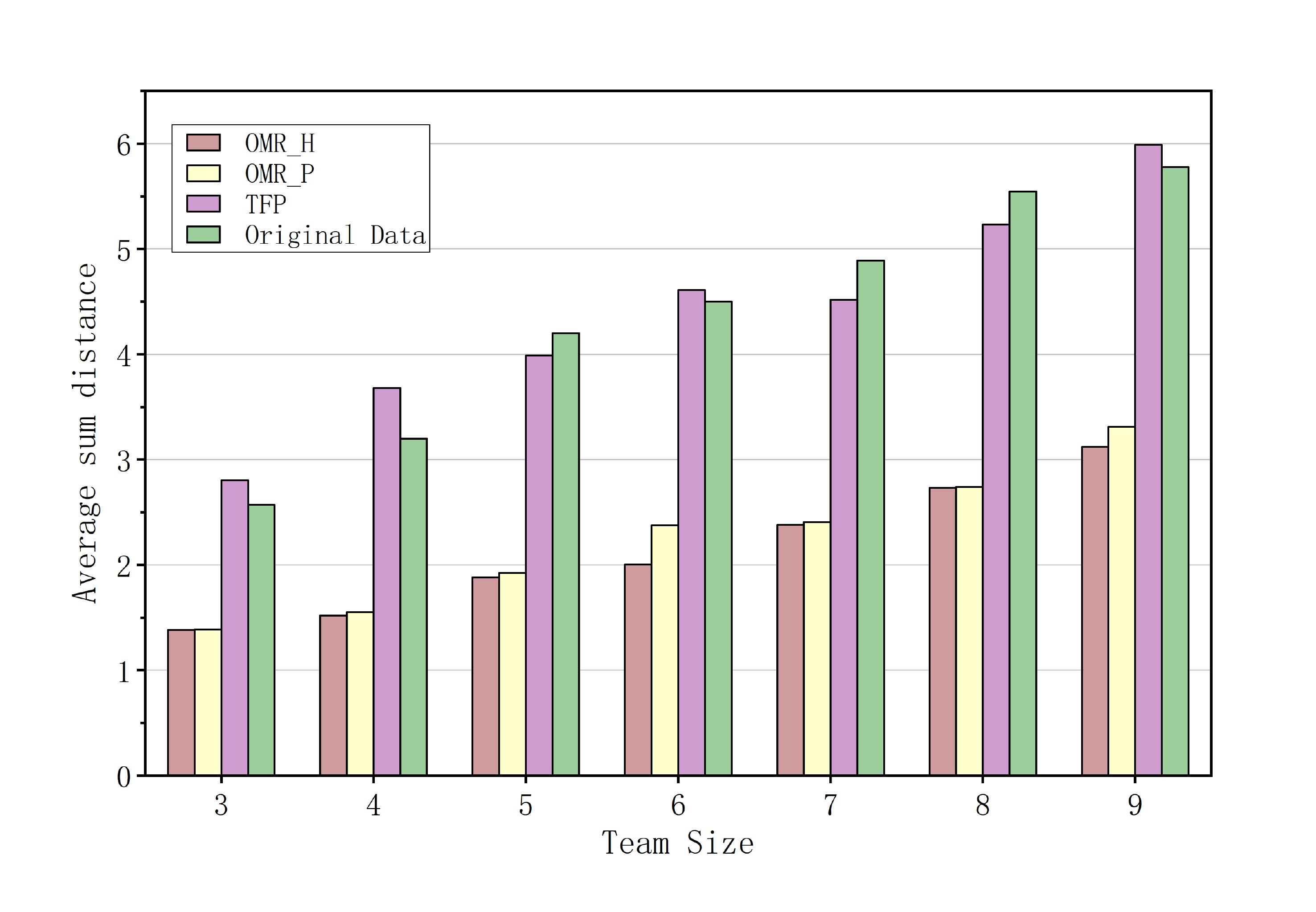}}
  \caption{Comparison results of sum distance in MAG dataset.}
  \label{fig:summag}
\end{figure}

The statistical data for the sum distance also indicates a similar conclusion, which can also been seen in Figure~\ref{fig:sum} and Figure~\ref{fig:summag}. In CiteSeerX, OMR\_H shortens the sum distance of original teams by 38\%, 52\%, 58\%, 51\%, 46\%, 38\%, 38\% for teams of size 3 to 9, respectively. These ratios are slightly lower if pairwise familiarity is employed. The sum distance for TFP is only reduced when team size is 4 and 9. The sum distance is more or less similar to the original team for the remaining team sizes. In MAG, we achieve similar experimental results. OMR\_H generally shortens the sum distance by 47\%, 20\%, 24\%, 23\%, 37\%, 25\%, 29\%. TFP shorten the sum distance when team size is 5, 7, and 8.

The above statistics indicate that the candidates OMR recommends generally lead to relative lower communication costs of new teams. The underlying reason is analyzed as follows. TFP approaches the issue of team replacement by finding a candidate who is similar to the team's outlier members, in the perspectives of skill and collaboration relationships. It simultaneously seeks a candidate who maintains intense collaboration with the remaining members. Therefore, the qualified candidates are either members of the original team or those who utilize more communication costs with the remaining members. However, OMR differs in that, the first impact factor considered is familiarity, which characteristically leads to lower communication costs.

Combining internal and external consumption, the proposed method OMR improves the optimization of academic teams upon the departure of outlier members.

\section{CONCLUSION}

Academic teams are an effective organizational structure that is used to improve scientific output through the implementation of academic activities. The efficacy of a team might be greatly hindered if members leave the team. To this end, a tenable solution to this issue has been proposed. Members who possess a high probability of leaving the team, are defined by evaluating the degree of familiarity among scholars. Pair-wise familiarity is defined on the basis of bi-partisan relationships and higher-order familiarity is defined using a multiplicity of relationships. The extent to which each member is an outlier is thereby quantified using such familiarity. Teams are then re-optimized after a threshold of outlier members have left. A random walk with graph kernel is employed to recommend a proper member by considering familiarity matching, skill matching, and structure matching. The CiteSeerX data set has been used for performance evaluation. The proposed approach which employs a higher-order familiarity outperforms other baseline methods.

\bibliographystyle{ACM-Reference-Format}
\bibliography{reference}


\end{document}